\documentclass{PoS}

\def\beq{\begin{equation}}
\def\eeq{\end{equation}}
\def\beqa{\begin{eqnarray}}
\def\eeqa{\end{eqnarray}}

\usepackage{epsfig}

\title{Theoretical results for top-quark cross sections and distributions}

\ShortTitle{Theoretical results for top-quark cross sections and distributions}

\author{\speaker{Nikolaos Kidonakis}\thanks{This material is based upon work supported by the National Science Foundation under Grant No. PHY 1519606.}\\
  Department of Physics, Kennesaw State University, Kennesaw, GA 30144, USA\\
        E-mail: \email{nkidonak@kennesaw.edu}}

\abstract{I present new results and updates for total cross sections and differential distributions in top-antitop pair and single-top production. Soft-gluon corrections are added to exact fixed-order results to provide the best predictions at approximate N$^3$LO for $t{\bar t}$ production and approximate NNLO for single-top production. Total cross sections and top-quark transverse-momentum and rapidity distributions are presented and compared with data at LHC and Tevatron energies. The cusp anomalous dimension at three and higher loops is also discussed.}

\FullConference{38th International Conference on High Energy Physics\\
		3-10 August 2016\\
		Chicago, USA}

\begin{document}

\section{Top-pair cross sections and differential distributions}

QCD corrections to $t{\bar t}$ production are quite substantial.
Soft-gluon corrections are a very important subset of the corrections 
and they approximate exact results very well. In fact, 
the higher-order results through NNLO have been very well predicted in the past
by the evaluation of the soft-gluon contributions \cite{NKnnlo}.
These soft corrections were calculated from resummation at NNLL accuracy for 
the double-differential cross section.
Approximate N$^3$LO (aN$^3$LO) predictions for cross sections were later 
derived \cite{NKttbar} by adding the third-order soft-gluon corrections.

The aN$^3$LO $t{\bar t}$ cross sections and top-quark differential 
distributions in $p_T$ and rapidity were calculated in \cite{NKttbar}, and 
numerical results were presented using MSTW2008 NNLO pdf \cite{MSTW2008}.
The cross sections increase somewhat if one uses the more recent 
MMHT2014 pdf \cite{MMHT2014}. For a top quark mass $m_t=173.3$ GeV 
the aN$^3$LO total cross section using MMHT2014 NNLO pdf is 
$826 {}^{+24}_{-16}{}^{+14}_{-18}$ pb at 13 TeV LHC energy 
and $975 {}^{+28}_{-19}{}^{+16}_{-20}$ pb at 14 TeV LHC energy, where 
the indicated uncertainties are from scale variation by a factor of two 
around $\mu=m_t$, and from the pdf at 68\% C.L. 

\begin{figure}
\begin{center}
\includegraphics[width=75mm]{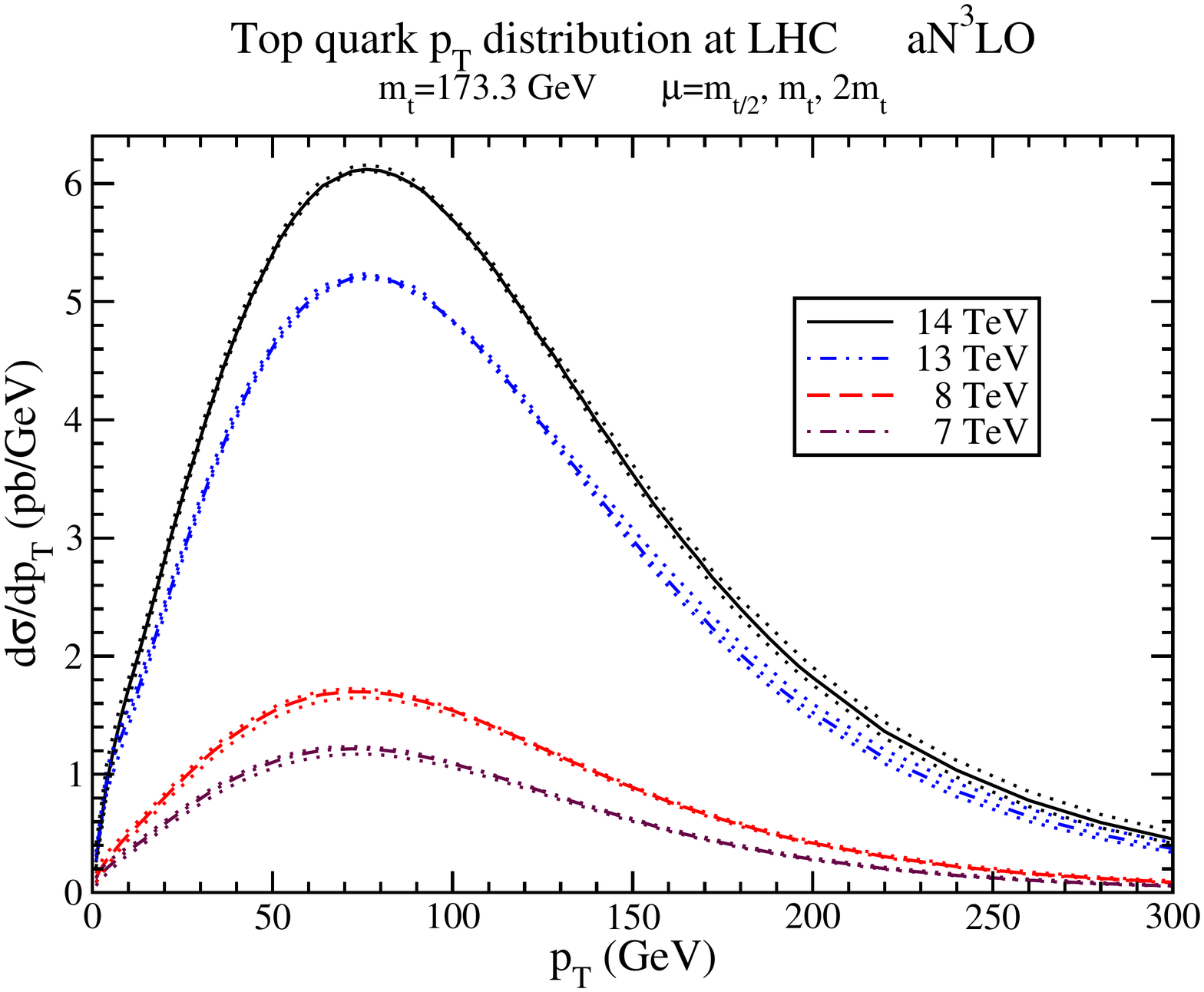}
\includegraphics[width=75mm]{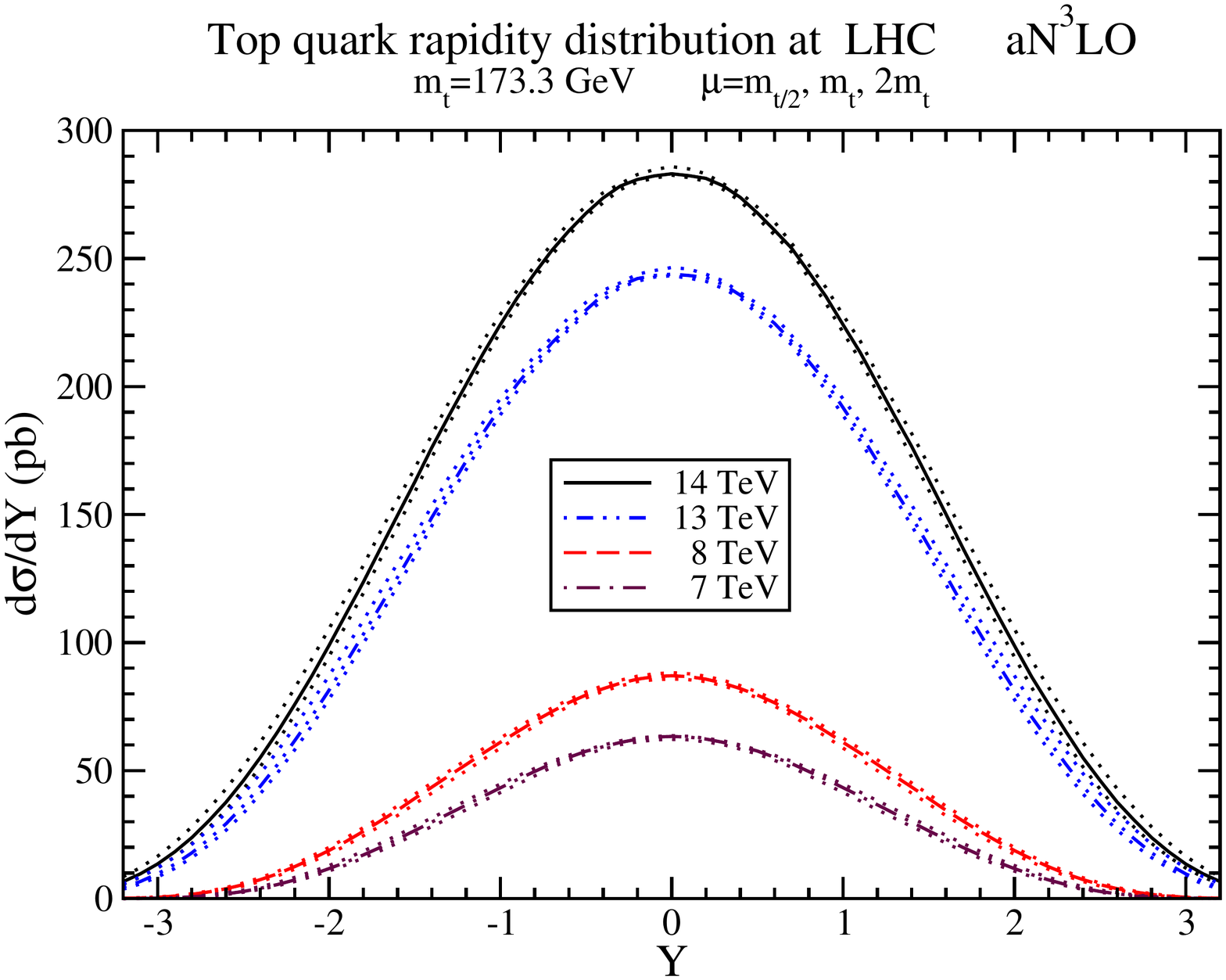}
\caption{Top-quark aN$^3$LO $p_T$ (left) and rapidity (right) distributions at LHC energies.}
\label{ptyaN3LO}
\end{center}
\end{figure}

In Fig. \ref{ptyaN3LO} we plot theoretical results, including scale variation, for the aN$^3$LO differential distributions in transverse momentum and rapidity of the top quark at 7, 8, 13, and 14 TeV LHC energies.

\begin{figure}
\begin{center}
\includegraphics[width=75mm]{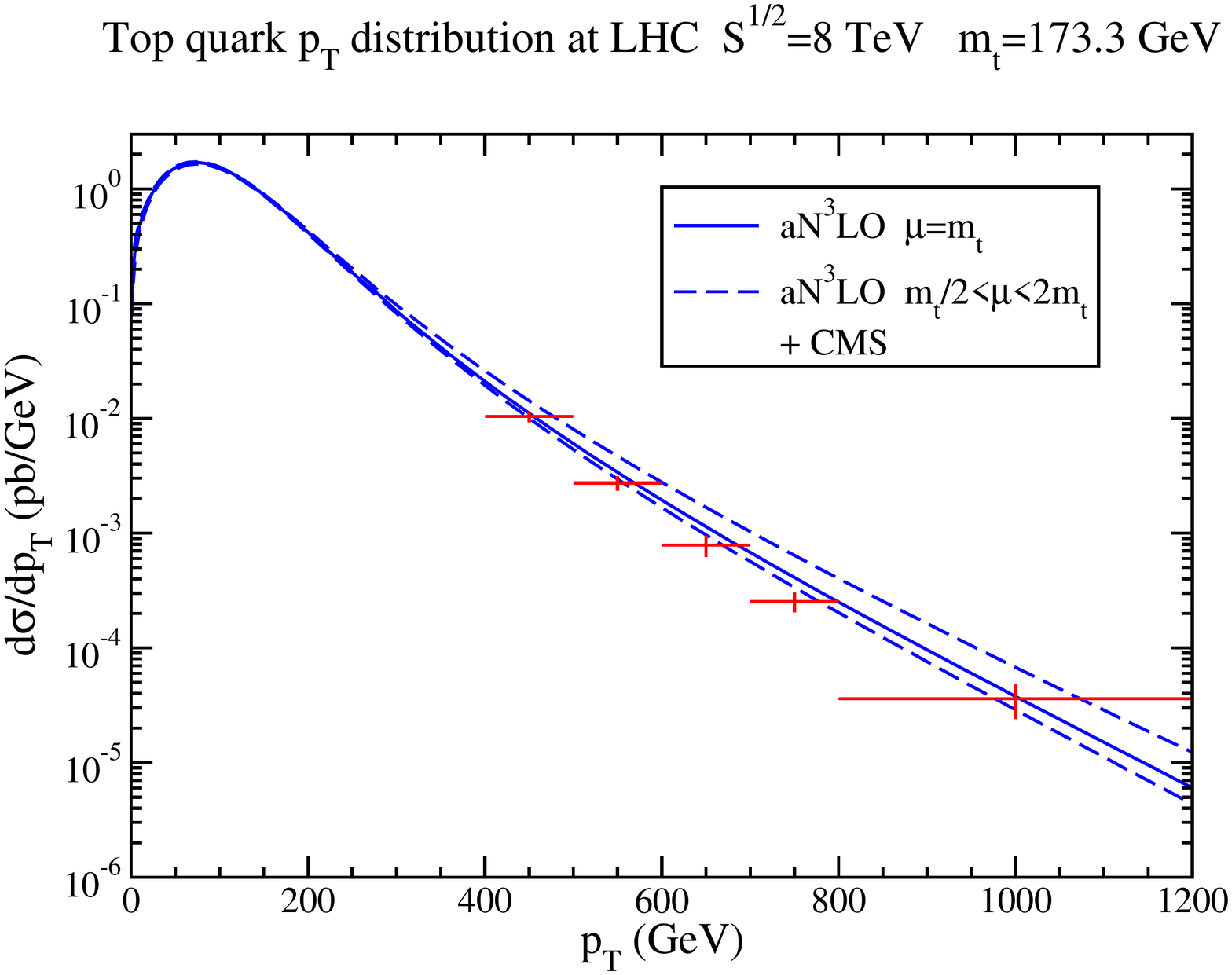}
\includegraphics[width=75mm]{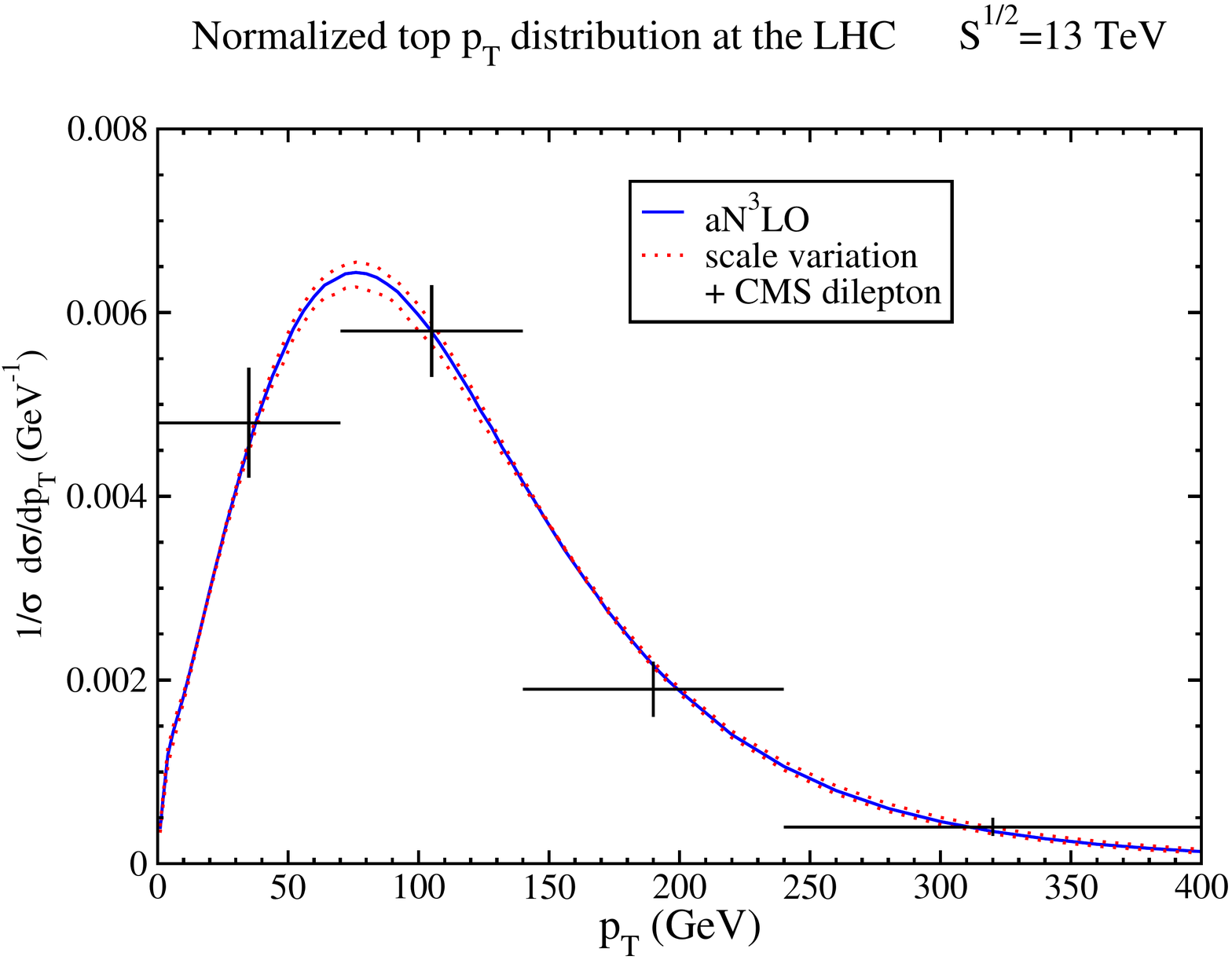}
\caption{Top-quark aN$^3$LO $p_T$ distributions at 8 TeV (left) and 13 TeV (right) LHC energies compared with CMS data \cite{CMSptlhc}.}
\label{ptlhc}
\end{center}
\end{figure}

In Fig. \ref{ptlhc} we plot theoretical results for the aN$^3$LO differential distributions in transverse momentum of the top quark at 8 and 13 TeV energies and compare with CMS data \cite{CMSptlhc}, finding excellent agreement between theoretical predictions and data in both cases. 

\begin{figure}
\begin{center}
\includegraphics[width=75mm]{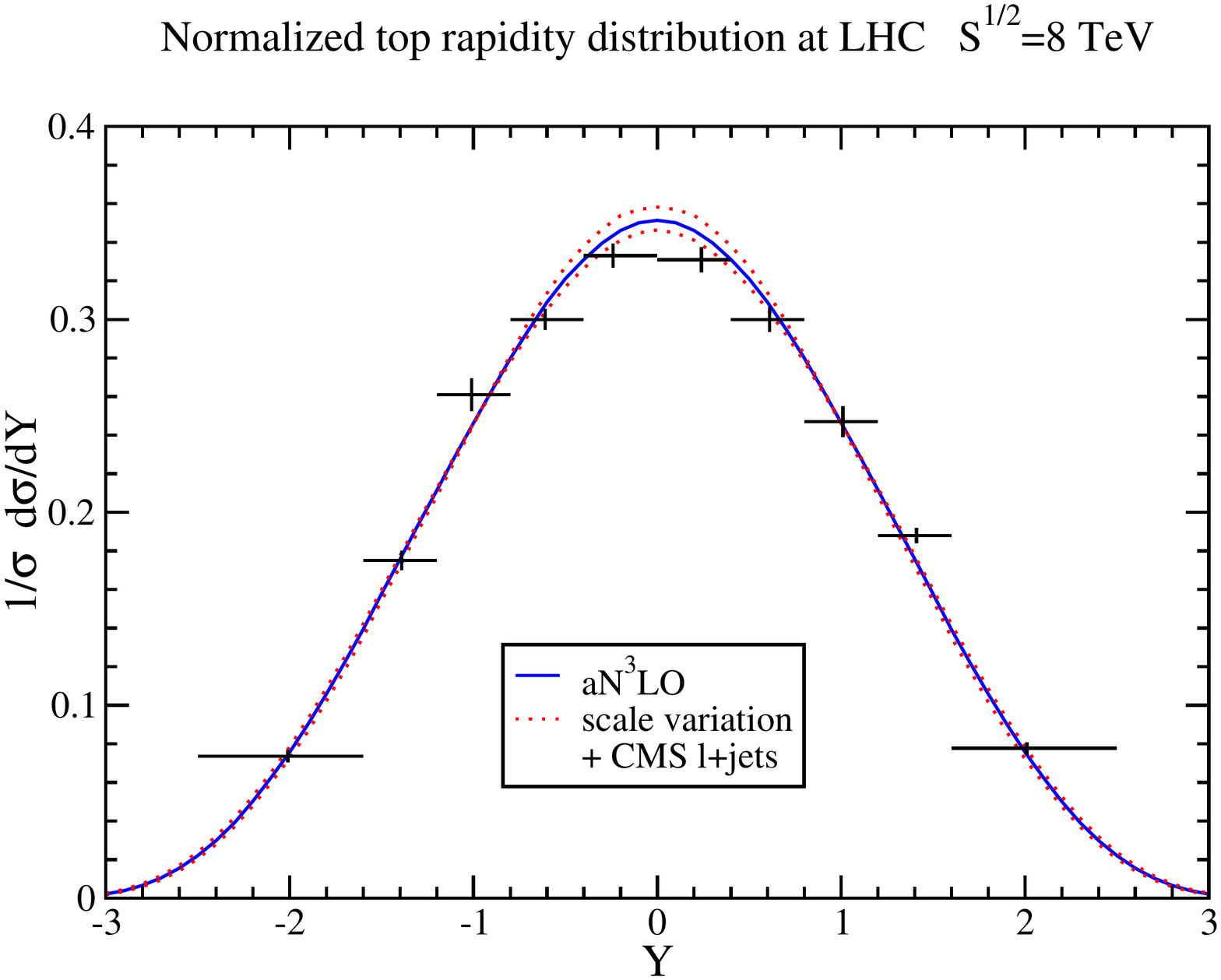}
\includegraphics[width=75mm]{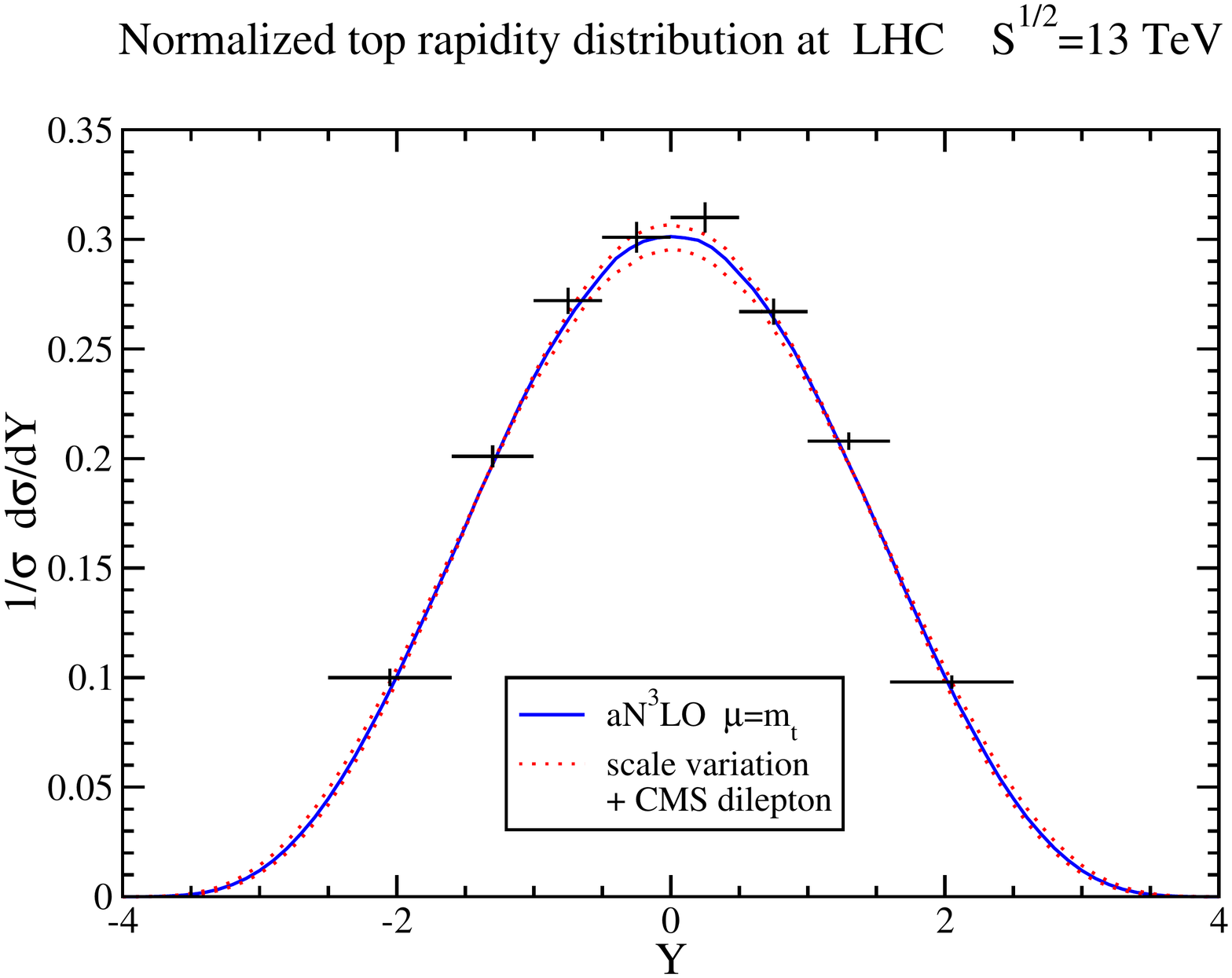}
\caption{Top-quark aN$^3$LO rapidity distributions at 8 TeV (left) and 13 TeV (right) LHC energies compared with CMS data \cite{CMSylhc}.}
\label{ylhc}
\end{center}
\end{figure}

In Fig. \ref{ylhc} we plot theoretical results for the aN$^3$LO normalized differential distributions in rapidity of the top quark and compare with CMS data at 8 and 13 TeV energies \cite{CMSylhc}, again finding excellent agreement between theory and data. 

Higher-order corrections are very sizable for the $t{\bar t}$ total cross 
sections and also for the differential distributions. Given the large 
contributions from higher 
orders and the relatively slow convergence of the perturbative series, it is 
clear that NNLO calculations are not enough; the inclusion of aN$^3$LO 
contributions is needed for truly precision top-quark physics.

\section{Single-top cross sections and differential distributions}

Single-top production processes include the $t$-channel via $qb \rightarrow q' t$ and ${\bar q} b \rightarrow {\bar q}' t$; the $s$-channel via $q{\bar q}' \rightarrow {\bar b} t$; and associated $tW$ production via $bg \rightarrow tW^-$.

\begin{figure}
\begin{center}
\includegraphics[width=88mm]{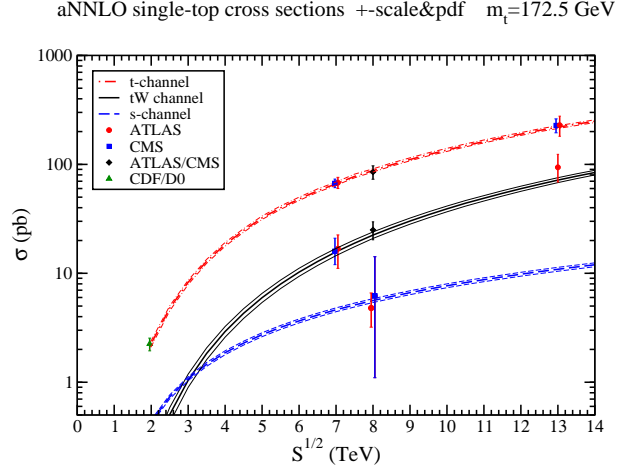} 
\caption{aNNLO single-top cross sections in $t$-channel, $s$-channel, and $tW$ processes compared with Tevatron \cite{tchtev} and LHC \cite{tchlhc,schlhc,tWlhc} data at various energies.}
\label{singletopplot}
\end{center}
\end{figure}

The aNNLO cross sections for all three processes were calculated in Ref. \cite{NKsingletop}, and results are plotted in Fig. \ref{singletopplot} with MSTW2008 pdf. We observe excellent agreement of theory with Tevatron and LHC data for the $t$-channel \cite{tchtev,tchlhc}, the $s$-channel \cite{schlhc}, and the $tW$ channel \cite{tWlhc}. 

The numbers increase a bit when MMHT2014 pdf are used. For the $t$-channel at the 13 TeV LHC using MMHT2014 NNLO pdf \cite{MMHT2014} and $m_t=173.3$ GeV, we find cross sections of $138 {}^{+3}_{-1} \pm 2$ pb  for the top and $83 {}^{+2}_{-1} \pm 1$ pb for the antitop, giving  $221 {}^{+5}_{-2} \pm 3$ pb for the sum. At 14 TeV we find $157 {}^{+4}_{-1} \pm 2$ pb for the top,  
$95 {}^{+2}_{-1} \pm 1$ pb for the antitop, and $252 {}^{+6}_{-2} \pm 3$ pb 
for the sum.

For the $s$-channel at 13 TeV LHC energy with MMHT2014 pdf 
we find cross sections of $7.15 \pm 0.13 {}^{+0.15}_{-0.17}$ pb 
for the top and $4.14 \pm 0.05 \pm 0.10$ pb for the antitop,  
giving $11.29 \pm 0.18 \pm 0.26$ pb for the sum.
At 14 TeV we find $7.83 \pm 0.14 \pm 0.18$ pb for the top,  
$4.60 \pm 0.05 \pm 0.11$ pb for the antitop, and  
$12.43 \pm 0.19 \pm 0.29$ pb for the sum.

For $tW^-$ production we find a cross section at 13 TeV of $36.3 \pm 0.9 \pm 0.9$  pb and at 14 TeV of $42.8 \pm 1.0 \pm 1.1$ pb with MMHT2014 pdf. The numbers are the same for 
${\bar t}W^+$ production.

\begin{figure}
\begin{center}
\includegraphics[width=75mm]{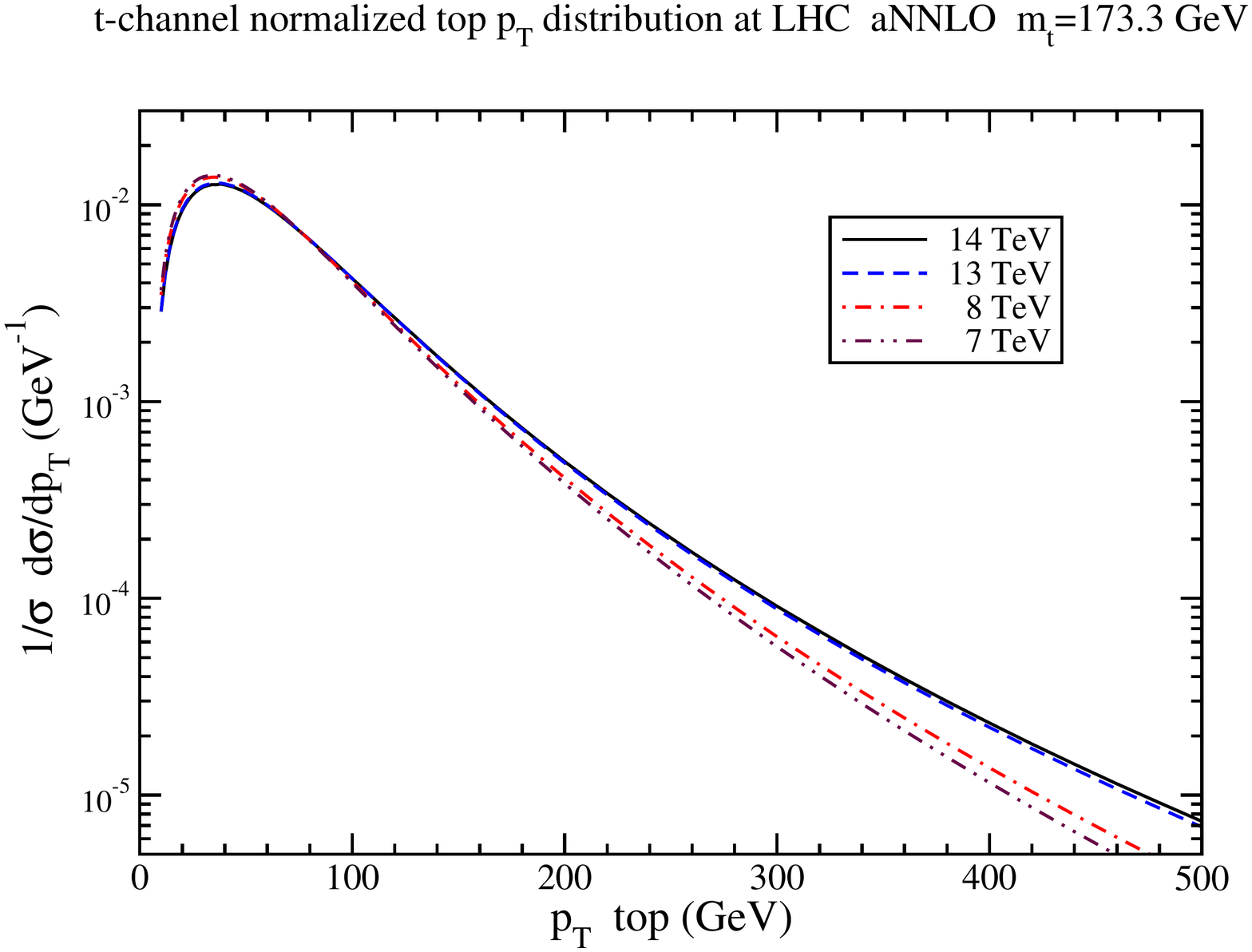}
\includegraphics[width=75mm]{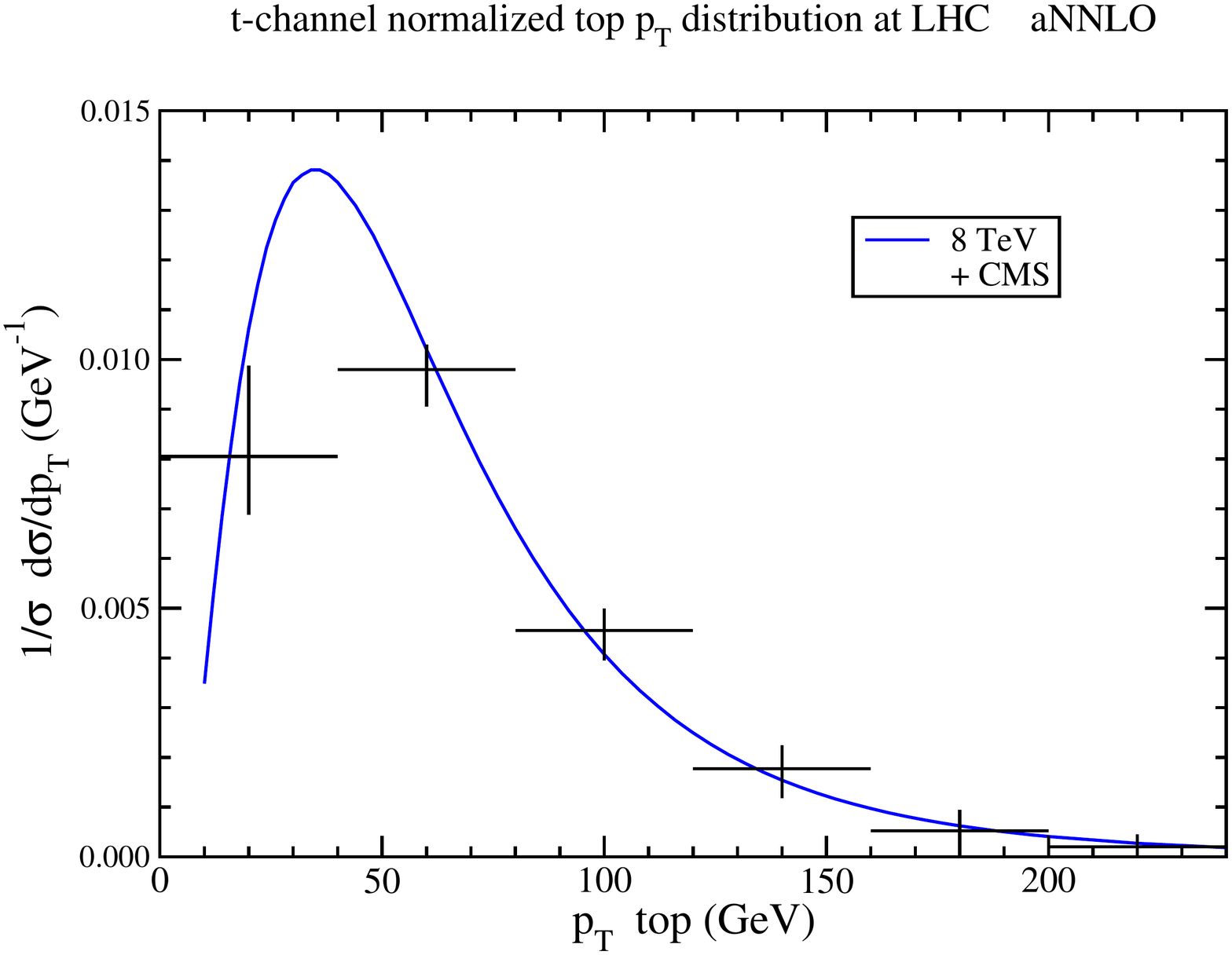} 
\caption{aNNLO top-quark normalized $p_T$ distributions in $t$-channel production at LHC energies (left) and comparison to CMS data \cite{CMStchpt8lhc} at 8 TeV energy (right).}
\label{pttchtopplot}
\end{center}
\end{figure}

In Fig. \ref{pttchtopplot} we plot the aNNLO top-quark normalized $p_T$ distributions in $t$-channel production at LHC energies (left plot) and compare with CMS data \cite{CMStchpt8lhc} at 8 TeV (right plot). 

\section{Cusp anomalous dimension}

\begin{figure}
\begin{center}
\includegraphics[width=75mm]{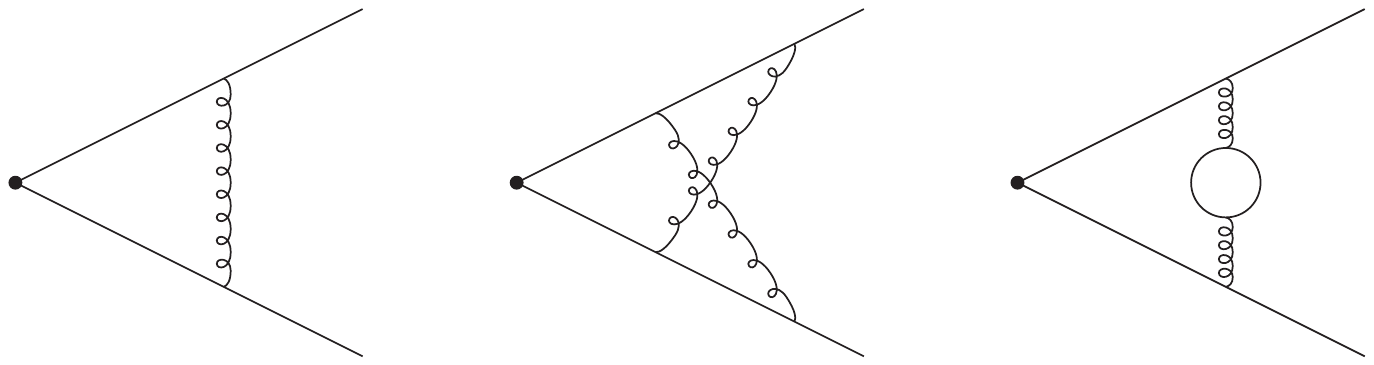} 
\hspace{-20mm}
\includegraphics[width=75mm]{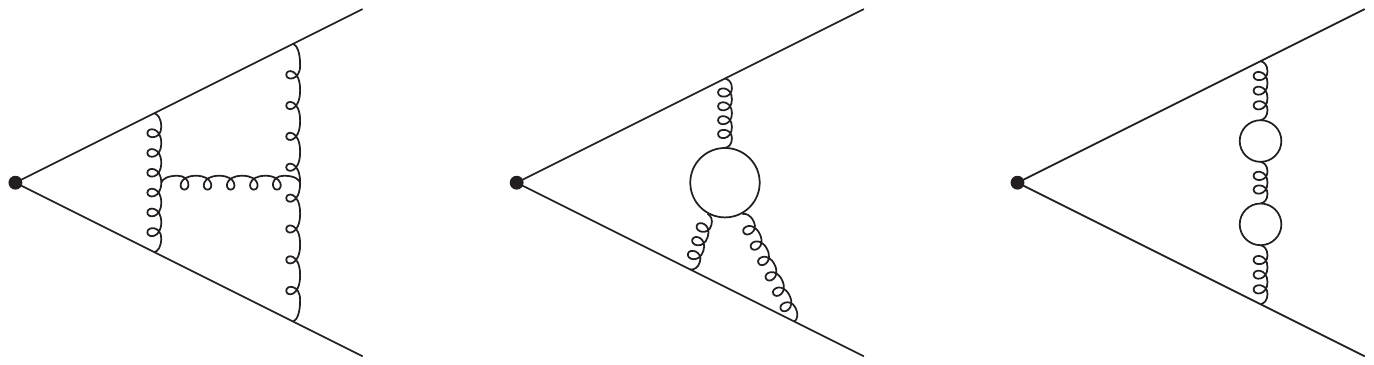} 
\vspace{-72mm}
\caption{Some of the one-loop, two-loop, and three-loop diagrams for the cusp anomalous dimension.}
\label{loops}
\end{center}
\end{figure}

The cusp anomalous dimension is an essential ingredient in higher-order calculations and resummations of soft-gluon contributions in perturbative cross sections. In particular, it is the simplest soft anomalous dimension, and a component of soft anomalous dimensions for more complicated color processes, such as $t{\bar t}$ production. Its perturbative expansion can be written as $\Gamma_{\rm cusp}=\sum_{n=1}^{\infty} (\alpha_s/\pi)^n \Gamma^{(n)}$ \cite{KR,NK2l,GHKM,NK3nl}. Some diagrams are shown in Fig. \ref{loops}.

\vspace{-5mm}

The cusp angle is $\theta=\cosh^{-1}(v_i\cdot v_j/\sqrt{v_i^2 v_j^2})$ where $v_i$ and $v_j$ are heavy-quark velocity vectors. We have
$\Gamma^{(1)}=C_F (\theta \coth\theta-1)$ and \cite{NK2l}
\beqa
\Gamma^{(2)}&=&\frac{K}{2} \, \Gamma^{(1)}
+\frac{1}{2}C_F C_A \left\{1+\zeta_2+\theta^2 
-\coth\theta\left[\zeta_2\theta+\theta^2
+\frac{\theta^3}{3}+{\rm Li}_2\left(1-e^{-2\theta}\right)\right] \right. 
\nonumber \\ && \hspace{30mm} \left.
{}+\coth^2\theta\left[-\zeta_3+\zeta_2\theta+\frac{\theta^3}{3}
+\theta \, {\rm Li}_2\left(e^{-2\theta}\right)
+{\rm Li}_3\left(e^{-2\theta}\right)\right] \right\} \, .
\nonumber
\eeqa

The three-loop result was first derived in \cite{GHKM}. In Ref. \cite{NK3nl} 
it was expressed in the form  
$$\Gamma^{(3)} = C^{(3)} +K^{'(3)} \Gamma^{(1)}
+K \left[\Gamma^{(2)}-\frac{K}{2}\Gamma^{(1)}\right]$$ 
where $C^{(3)}$ was given in terms of ordinary polylogarithms.

From the structure of $\Gamma_{\rm cusp}$ through three loops, an $n$-loop conjecture was put forth in \cite{NK3nl}:
$$
\Gamma^{(n)}=\sum_{k=1}^n \frac{(n-1)!}{(k-1)! \, (n-k)!} \, K^{'(k)} \, 
C^{(n-k+1)} \, .
$$

\begin{figure}
\begin{center}
\includegraphics[width=75mm]{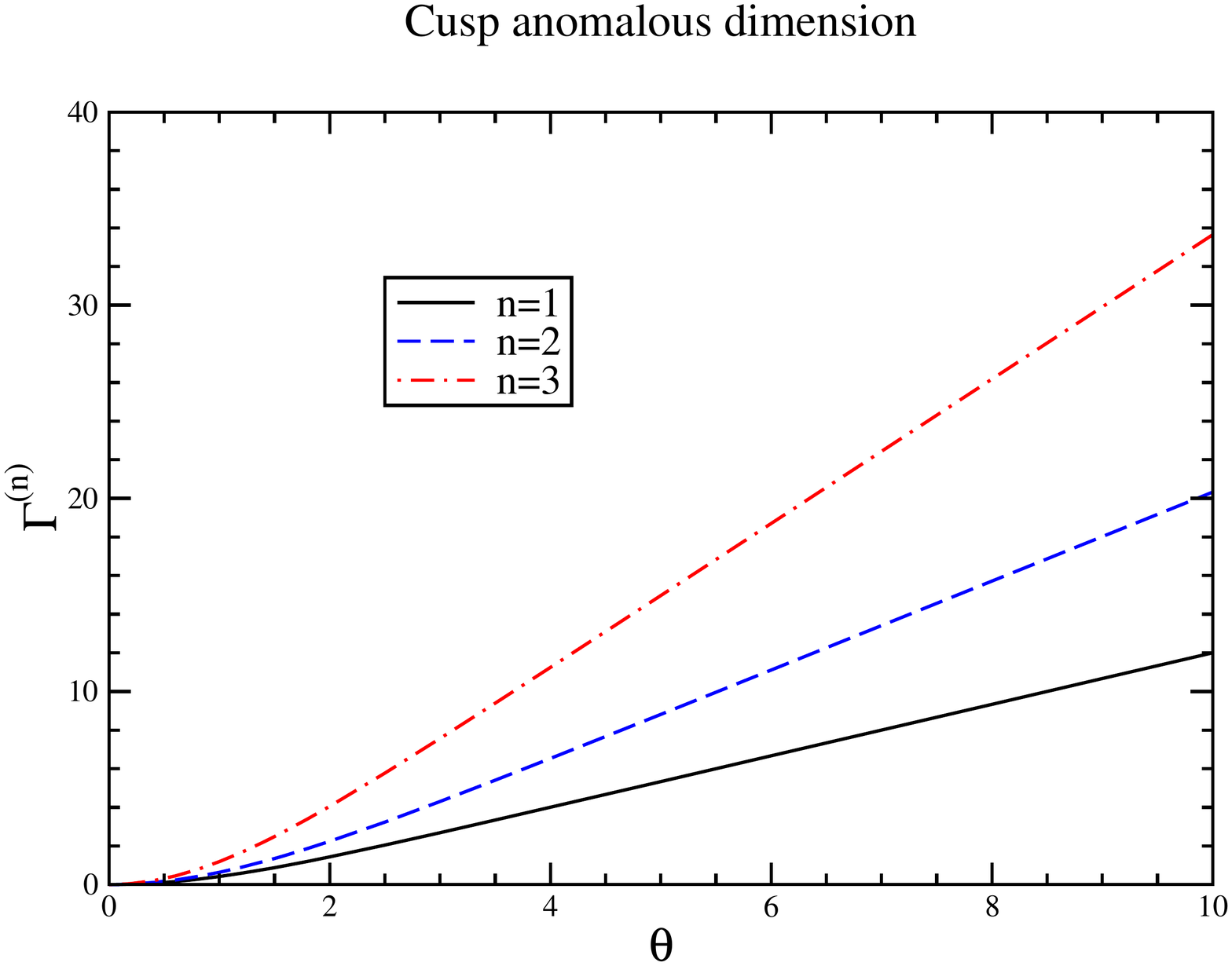} 
\includegraphics[width=75mm]{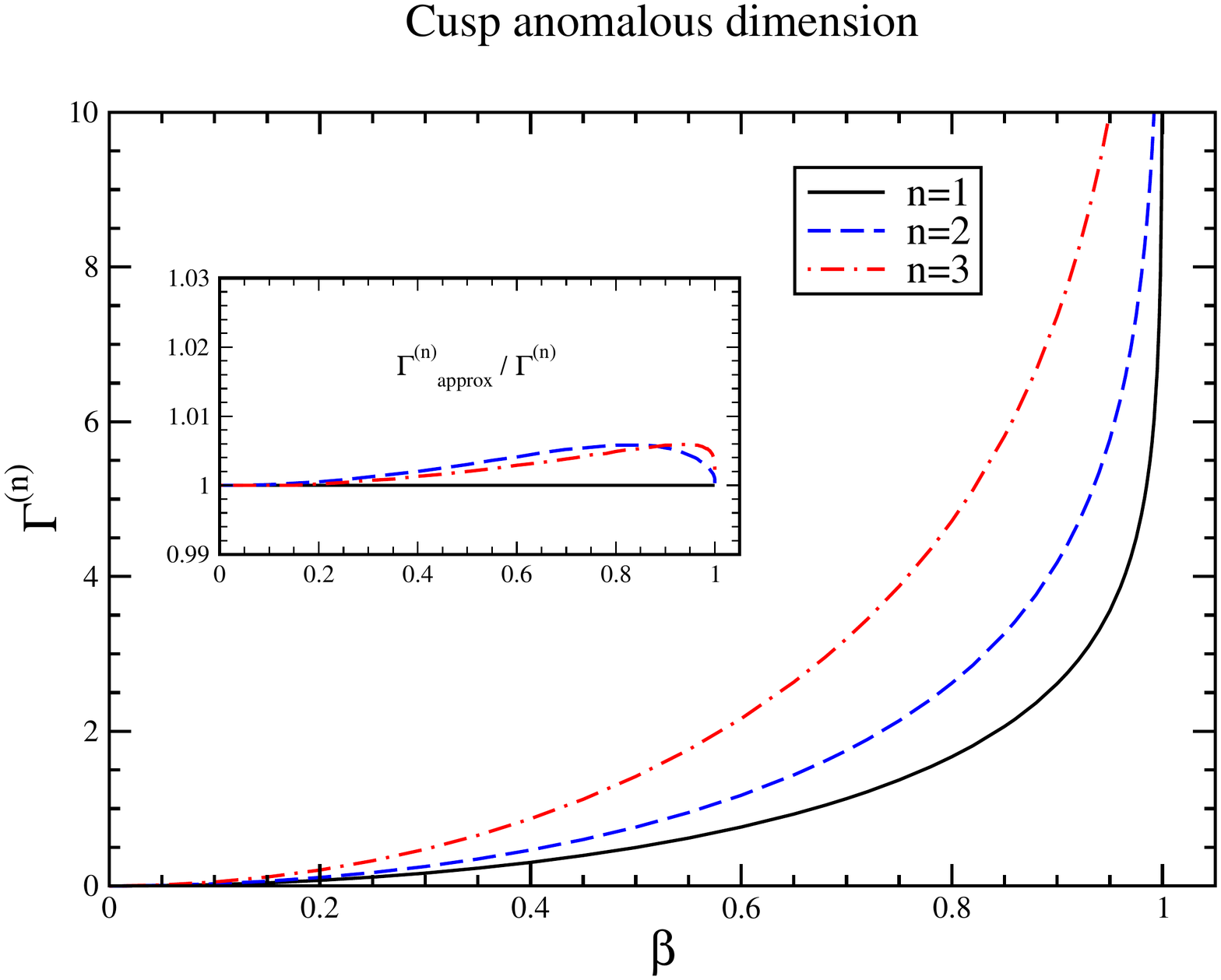} 
\caption{The cusp anomalous dimension at one, two, and three loops.}
\label{gammacusp}
\end{center}
\end{figure}

Simple yet excellent numerical approximations to the exact results can be derived.  We find, for $n_f=5$ and with $\beta=\tanh(\theta/2)$, the simple formulas \cite{NK3nl}
$$
\Gamma^{(2)}_{\rm approx}(\beta)=
-0.38649 \, \beta^2+1.72704 \; \Gamma^{(1)}(\beta) \, ; \quad 
\Gamma^{(3)}_{\rm approx}(\beta)=
0.09221 \, \beta^2+2.80322 \; \Gamma^{(1)}(\beta) \, .
$$

In Fig. \ref{gammacusp} we plot the cusp anomalous dimension and its approximations through third order.

\end{document}